\documentclass[prl,twocolumn,floatfix,showpacs]{revtex4-1}

\usepackage{graphicx}
\usepackage[latin1]{inputenc}
\usepackage{dcolumn}
\usepackage{bm}

\begin{document}


\title{Magnetoresistance Induced by Rare Strong Scatterers in a High-Mobility 2DEG}

\author{L. Bockhorn$^1$}
\email{bockhorn@nano.uni-hannover.de}
\author{I. V. Gornyi$^2$}
\author{D. Schuh$^3$}
\author{C. Reichl$^4$}
\author{W. Wegscheider$^4$}
\author{R. J. Haug$^1$}

\affiliation{$^1$Institut f\"ur Festk\"orperphysik, Leibniz Universit\"at Hannover, D-30167 Hannover, Germany\\
$^2$Institut f\"ur Nanotechnologie, Karlsruhe Institute of Technology, D-76021 Karlsruhe, Germany\\
and A.F.~Ioffe Physico-Technical Institute,
 194021 St.~Petersburg, Russia\\
$^3$Institut f\"ur Experimentelle und Angewandte Physik, Universit\"at Regensburg, D-93053 Regensburg, Germany\\
$^4$Laboratorium f\"ur Festk\"orperphysik, ETH Z\"urich, CH-8093 Z\"urich, Switzerland}

\date{\today}

\begin{abstract}
We observe a strong negative magnetoresistance at non-quantizing magnetic fields in a high-mobility two-dimensional electron gas (2DEG). This strong negative magnetoresistance consists of a narrow peak around zero magnetic field and a huge magnetoresistance at larger fields. The peak shows parabolic magnetic field dependence and is attributed to the interplay of smooth disorder and rare strong scatterers. We identify the rare strong scatterers as macroscopic defects in the material and determine their density from the peak curvature.
\end{abstract}


\maketitle

The quality and the mobility of two-dimensional electron gases~(2DEG) have improved continuously since the first observation of the fractional Quantum Hall effect~(FQHE)\cite{Tsui1982, Laughlin1983}. This improvement has not only allowed the observation of new effects but also led to emergence of new questions. One problem is the characterization of the sample quality which is usually reflected by the electron mobility~$\mu_e$ determined from the resistivity at zero magnetic field. Sometimes, FQHE features are observed for a variety of mobilities but not in the highest mobility samples. It follows that the electron mobility alone cannot serve as a reliable indication for the quality of high-mobility samples. Therefore, one has to look for more specific effects to characterize the sample quality. The quality of high-mobility samples depends on various types of disorder. The remote donors are normally assumed to be the main source of disorder. Other sources like e.~g. the scattering on interface roughness and on residual impurities in the quantum well become important with increased spacer width. Here, we consider macroscopic defects in the sample as an additional source of disorder. These so called oval defects are seen as randomly distributed defects on the material surface. In this letter we show that a negative magnetoresistance around zero magnetic field is induced by such macroscopic defects.

A recent work reported on a strong negative magnetoresistance at non-quantizing magnetic fields~\cite{Bockhorn2011} which consists of a peak around zero magnetic field and a huge negative magnetoresistance at larger magnetic fields. In Ref.~\cite{Bockhorn2011} the focus was on the huge magnetoresistance while the peak was considered as a geometry effect. In the present paper, we analyze the peak around zero magnetic field in more detail and observe that the peak is induced by the interplay of two types of disorder, smooth disorder due to remote ionized impurities and rare strong scatterers due to the presence of macroscopic defects. The density of macroscopic defects is deduced from the peak considering the theory developed in Ref.~\cite{Mirlin2001, Polyakov2001}.

\begin{figure}
   \centering
   \includegraphics{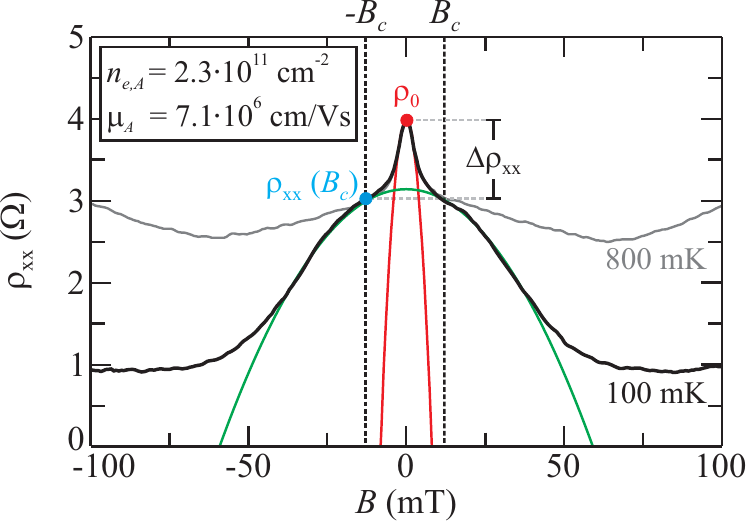}
   \caption{\label{Strong} The longitudinal resistivity~$\rho_{xx}$ vs. magnetic field~$B$ for two temperatures~$T$. The strong negative magnetoresistance is divided into two sections fitted by parabolic magnetic field dependence. The huge magnetoresistance (green parabola) depends strongly on the temperature, while the narrow peak around zero magnetic field (red parabola) is left unchanged.}
\end{figure}

Various ungated and gated samples of two different materials (Sample A and B) were used for the magnetotransport measurements. The behavior of the strong negative magnetoresistance was always similar. The 2DEG of both materials is realized in a GaAs/AlGaAs quantum well grown by molecular-beam epitaxy. The quantum well has a width of 30$\,$nm and is Si-doped from both sides with spacers of $70\,$nm. Only the electron density and the electron mobility is slightly different for both materials (\mbox{$n_{e,A}\approx3.1\cdot10^{11}\,$cm$^{-2}$} and \mbox{$n_{e,B}\approx3.3\cdot10^{11}\,$cm$^{-2}$}, respectively \mbox{$\mu_A\approx11.9\cdot10^{6}\,$cm$^{2}$/Vs} and \mbox{$\mu_B\approx10.9\cdot10^{6}\,$cm$^{2}$/Vs}). The specimens are Hall bars with a total length of 1.2$\,$mm, a width of $w=200\,\mu$m and a potential probe spacing of $l=300\,\mu$m. The Hall bars were defined by photolithography and wet etching. The magnetotransport measurements were performed in a dilution refrigerator with a base temperature of 20$\,$mK. All measurements were carried out by using low-frequency (13$\,$Hz) lock-in technique.

Figure~\ref{Strong} shows a typical measurement of the strong negative magnetoresistance at $|B|<\,$100$\,$mT for two different temperatures. We divide the negative magnetoresistance into two distinct sections with parabolic magnetic field dependences. The huge magnetoresistance at larger magnetic fields (green parabola) depends strongly on the temperature, while the narrow peak around zero magnetic field (red parabola) is left unchanged for low temperatures~\cite{Bockhorn2011, Hatke2012} which is a sign for the absence of weak localization. The crossover between the peak and the huge magnetoresistance is seen as a shoulder in the longitudinal resistance around $B_c=\pm12\,$mT. 

\begin{figure}
   \centering
   \includegraphics{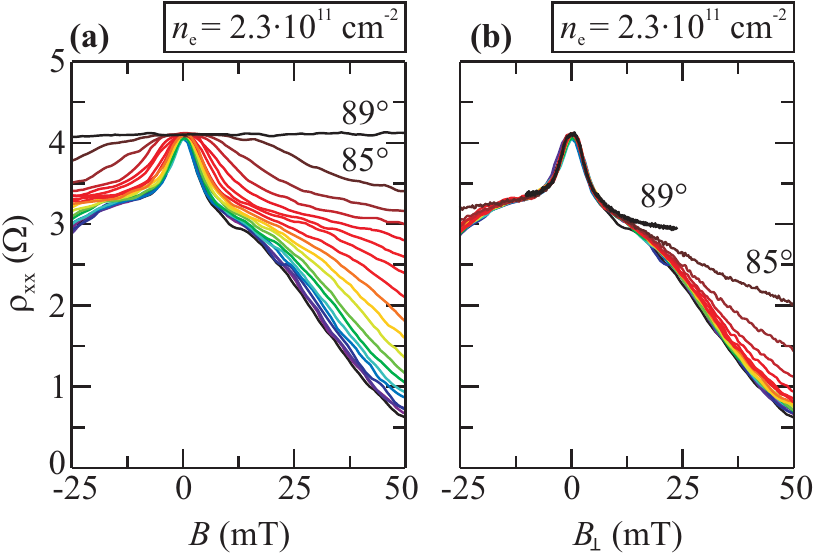}
   \caption{\label{Winkel}\textbf{(a)}~The strong negative magnetoresistance vs. total magnetic field~$B$ for different angles. The angle is increased from 0° till 90° in steps of 5°.~\textbf{(b)}~The strong negative magnetoresistance vs. perpendicular magnetic field~$B_\bot$ for the same angles. The huge magnetoresistance shows a tilt angle dependence, while the peak is left unchanged.}
\end{figure}

In order to understand the nature of the different types of magnetoresistances, we examine the effect of an in-plane magnetic field component on the strong negative magnetoresistance. The in-plane magnetic field is introduced by tilting the sample with respect to the magnet axis. In Fig.~\ref{Winkel}~(a) the longitudinal resistivity~$\rho_{xx}$ is shown vs. total magnetic field~$B$ for different tilt angles. The tilt angle is increased in steps of 5° from 0° to 90°. The width of the peak and the width of the huge magnetoresistance increase with tilt angle. To test the two-dimensionality of the observed effect Figure~\ref{Winkel}~(b) shows the longitudinal resistance~$\rho_{xx}$ vs. perpendicular magnetic field~$B_\bot$ for the corresponding tilt angles. The curvature of the huge magnetoresistance is constant till 60°. Above 60° the curvature decreases by increasing the tilt angles as also observed in Ref.~\cite{Hatke2012}. Therefore, the huge magnetoresistance shows a tilt angle dependence which hints towards an influence of the three-dimensionality of the sample material. More specifically, in the presence of a parallel component of magnetic field the wavefunctions of electrons are shifted to one side of the quantum well, so that the scattering of the electrons with the lattice increases. Consequently, the scattering rate rises and the huge magnetoresistance vanishes~\cite{Inarrea2014}.

In contrast the peak around zero magnetic field is left unchanged for all tilt angles as function of perpendicular magnetic field. The tilt angle independence of the peak means that it is a purely two-dimensional effect.

\begin{figure}
   \centering
   \includegraphics{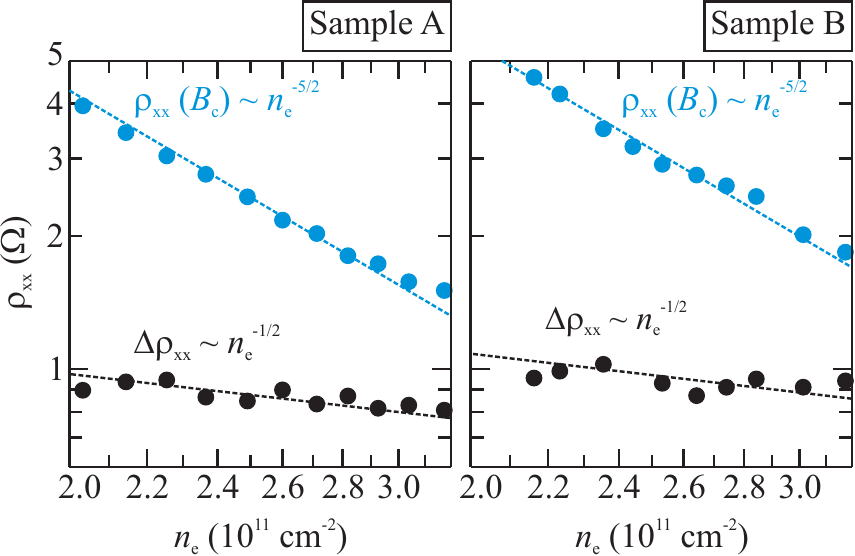}
   \caption{\label{Peak} The value of~$\rho_{xx}(B_c)$ and the height of the peak~$\Delta\rho_{xx}$ are shown vs. electron density~$n_e$ using a log-log scale. Both $\rho_{xx}(B_c)$ and $\Delta\rho_{xx}$ decrease with increasing electron density.}
\end{figure}

In the following part we analyze the behavior of the peak in more detail. The peak is characterized by two quantities, the height of the peak and its curvature. The height of the peak $\Delta\rho_{xx}=\rho_0-\rho_{xx}(B_c)$ is given by the difference between the resistivity $\rho_0$ at zero magnetic field and the value of the shoulder~$\rho_{xx}(B_c)$. The curvature of the peak is determined by fitting a parabola to the experimental data. Figure~\ref{Peak} shows the value of $\rho_{xx}(B_c)$ and the corresponding height of the peak~$\Delta\rho_{xx}$ vs. electron density~$n_e$ for both materials. Both $\rho_{xx}(B_c)$ and the height of the peak~$\Delta\rho_{xx}$ decrease with increasing electron density. Note that the shoulder in the longitudinal resistivity shows a stronger power-law dependence on the electron density, $\rho_{xx}(B_c)\propto\,n_e^{-5/2},$ than the height of the peak, $\Delta\rho_{xx}\propto n_e^{-1/2}$.

The dependence of \mbox{$\rho_{xx}(B_c)\propto\,n_e^{-5/2}$} on electron density is characteristic for scattering by smooth disorder. In a 2DEG with a smooth random potential the corresponding transport scattering time is given by
\begin{equation}
	\tau_L^{-1}\sim \frac{v_F}{d}\left( \frac{U}{E_F}\right) ^2
	\label{eq:tauL}
\end{equation}
with $d$ the correlation radius and $U$ the amplitude of the potential, $v_F$ the Fermi velocity, and $E_F$ the Fermi energy. The subscript~$L$ in eq.~(\ref{eq:tauL}) emphasizes the long-range character
of this type of disorder in contrast to the short-range disorder. For the model of smooth disorder created by remote donors the correlation radius~$d$ of the smooth disorder is determined by the spacer width i.~e.~$d\geq70\,$nm. The resulting transport scattering time \mbox{$\tau_L^{-1}=\pi\,n_{i}\,\hbar/[4\,m^*\left(k_F\,d\right)^3]$} depends on electron density as $n_e^{-3/2}$~\cite{Dmitriev2012, footnote-tauL}. Here $n_i$ is the effective 2D density of donors, $k_F$ is the Fermi wavevector and $m^*$ is the effective mass. If the resistivity is dominated by smooth disorder, one expects the following dependence on electron density
\begin{equation}
	\rho_{xx}=\frac{m^*}{e^2n_e\tau_L}\propto n_e^{-5/2}.
	\label{eq:rho}
\end{equation}
The same dependence is observed in Fig.~\ref{Peak} for the value of the shoulder $\rho_{xx}(B_c)$ in the longitudinal resistivity. Therefore, it is natural to assume that the main scattering mechanism governing the resistivity in fields higher than $B_c$ is provided by smooth disorder. To verify this statement we compare the quantum relaxation time~$\tau_q$ with the transport scattering time~$\tau_L$. The quantum relaxation time~$\tau_q$ is calculated from the magnitude of the SdH-oscillations following \mbox{Coleridge~\textit{et al.}~\cite{Coleridge1989}}. For an electron density of $n_e=3.2\cdot10^{11}\,$cm$^{-2}$ the transport scattering time deduced at $B_c$ \mbox{($\tau_L=4.9\cdot10^{-10}\,$s)} is much larger than the quantum relaxation time~\mbox{$\tau_q=1.8\cdot10^{-12}\,$s}. The large ratio of $\tau_L/\tau_q\sim270$ shows the dominance of small-angle scattering at remote ionized impurities and is close to the theoretically expected ratio of $(2\,k_F\,d)^2$. This implies that the main scattering mechanism in our samples is not due to background (short-ranged) impurities \cite{footnote-background}. On the other hand, the height of the peak~$\Delta\rho_{xx}$ does not scale as $n_e^{-5/2}$ and hence an additional type of disorder has to play a role.

In Ref.~\cite{Bockhorn2011} it was assumed that the peak around zero magnetic field is given by scattering at the edges in the ballistic regime, similar to the quenching of the Hall effect~\cite{Roukes1987, Thornton1989}. Our analysis of the strong negative magnetoresistance for different length-to-width ratios shows that the peak is independent of the geometry~\cite{Bockhorn2013}, in contrast to the recent observation~\cite{Mani2013} for similar samples. In particular, we would also expect the peak to be larger for higher electron densities if it would depend on the ratio between classical cyclotron orbit and geometry. Instead, the combination of our observations can be consistently described within the model of the interplay of two types of disorder as discussed in Ref.~\cite{Mirlin2001, Polyakov2001}.

Specifically, Ref.~\cite{Mirlin2001} calculated a negative magnetoresistance induced by an interplay of smooth disorder and rare strong scatterers. The combination of both types of disorder induces a novel mechanism of negative magnetoresistance due to memory effects leading to a peak around zero magnetic field. This negative magnetoresistance is followed by a saturation of the longitudinal resistivity~$\rho_{xx}(B)$ at a value determined by smooth disorder. This is in agreement with our observations of the scaling of $\rho_{xx}(B_c)$ above. In stronger magnetic fields the effect of rare strong scatterers is negligible. Other mechanisms of magnetoresistance become more efficient with further increased magnetic field, so that instead of the saturation one observes a dependence of $\rho_{xx}$ on magnetic field characterized by different scales.

On a quantitative level, the interplay of a long-range smooth random potential and strong scatterers is governed by the ratio of the corresponding mean free paths. The mean free path due to scattering by smooth disorder is $\ell_L=v_{F}\tau_{L}$ with the transport relaxation time $\tau_L=m^*/(e^2 n_e \rho_{xx}(B_c))$ and $v_{F}=\hbar k_F/m^*$ the Fermi velocity. The mean free path due to the randomly distributed strong scatterers is $\ell_S=v_{F}\tau_{S}\sim\,n_S\cdot\,a_S$, where $\tau_S$ is the transport scattering time due to scattering by strong scatterers, $n_S$ is the density of the rare strong scatterers and $a_S$ is the radius of the strong scatterers. We assume $\tau_L\sim\tau_S$ in the situation of high-mobility samples. Within the model of Ref.~\cite{Mirlin2001}, there is a crossover from $\rho_0=(m^*/e^2\,n_e)(\tau^{-1}_L+\tau^{-1}_S)$ to $\rho_{xx}(B_{c})=m^*/(e^2\,n_e\,\tau_L$) which takes places around the 'percolation threshold' $B_{c}$. Below the percolation threshold~$B_c$ electrons move in rosettelike trajectories around the strong scatterers. For larger magnetic fields~$B>B_c$ this rosettelike movement of the electrons is precluded by other scattering events. In Fig.~\ref{Strong} we observe a saturation of the longitudinal resistivity around $\pm12\,$mT, this shoulder marks the percolation threshold $B_{c}$. 

In our measurements the height of the peak~$\Delta\rho_{xx}$ is given by $\rho_0-\rho_{xx}(B_c)=m/(e^2\,n_e\,\tau_S)$. From Fig.~\ref{Peak} we can conclude that $\tau_S\sim\,n_e^{-1/2}$. We find that $\tau_L$ and $\tau_S$ have the same order for the considered range of electron densities. This is clearly seen from the comparison of the magnitudes of the peak and huge magnetoresistance in Fig.~\ref{Peak}, suggesting roughly \mbox{$\tau_S\sim 3\,\tau_L$ to $5\,\tau_L$}. The observed dependence on electron densities of $\tau_L\propto n_e^{3/2}$ and $\tau_S\propto n_e^{-1/2}$ confirms the different natures of the narrow peak and the huge magnetoresistance~\cite{Bockhorn2011}. The peak is then expressed by
\begin{equation}
	\frac{\rho_{xx}}{\rho_{0}}=1-\frac{\omega^{2}_{c}}{2\,\pi\,n_S\,v_F^2}\,f(x),
	\label{eq:1}
\end{equation}
where \mbox{$\omega_{c}=eB/m^*$} is the cyclotron frequency,
\begin{equation}
f(x)=\frac{2}{x+1}\int_0^\infty\!\!\,dq\,\frac{q\ J_1^2(q)}{x\,q^2+2[1-J_0^2(q)]}
\end{equation}
with \mbox{$x=\tau_S/\tau_L$}, and $J_{0,1}(q)$ Bessel functions. It is a generalization of the result which was derived in Ref.~\cite{Mirlin2001} in the limit $x\ll 1$.

The mixed-disorder model~\cite{Mirlin2001} was also used by Dai et al.~\cite{Dai2010} to describe a negative magnetoresistance. In Ref.~\cite{Dai2010} no distinct shoulder (seen as a two-scale negative magnetoresistance) was observed. Unfortunately, Ref.~\cite{Dai2010} did not report on the temperature dependence of the magnetoresistance which in our case serves as an important tool to distinguish between different mechanisms.

\begin{figure}
   \centering
   \includegraphics{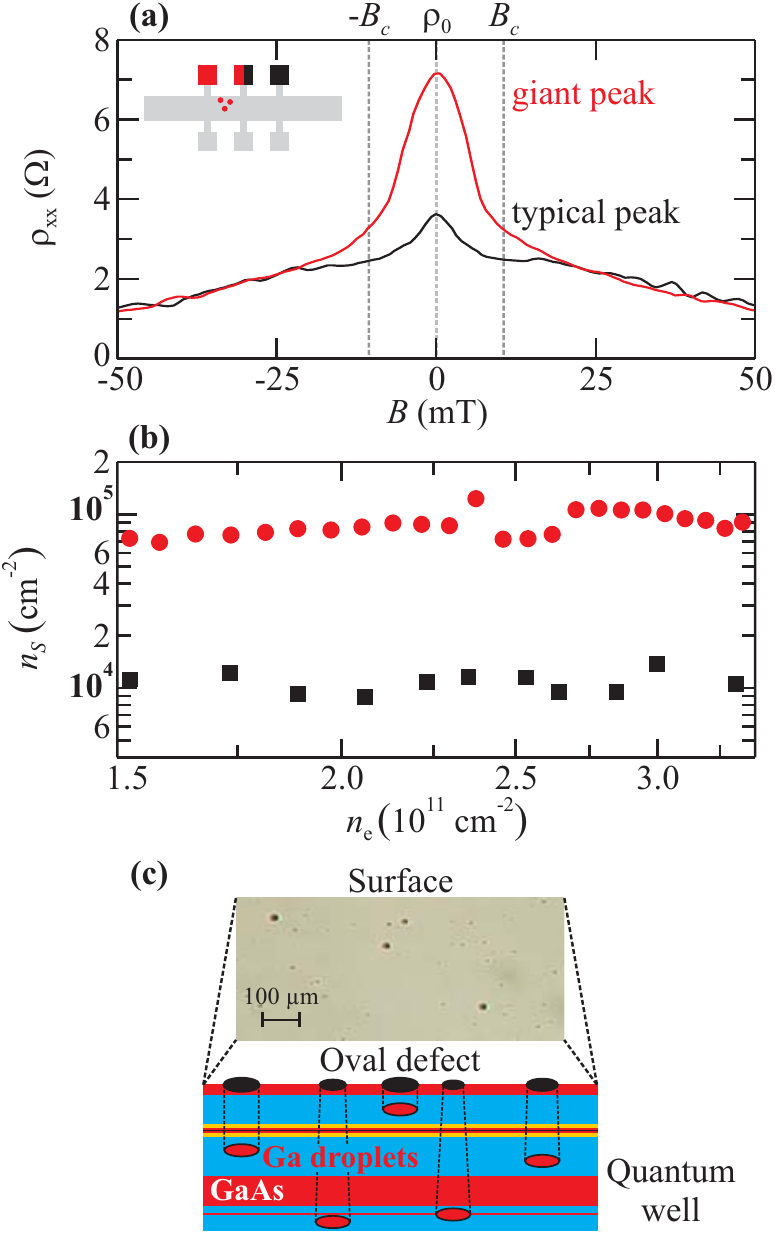}
   \caption{\label{Streuer} (a)~The longitudinal resistivity~$\rho_{xx}$ vs. magnetic field~$B$ for two different ohmic contacts at \mbox{$n_e=2.7\cdot10^{11}\,$cm$^{-2}$}. We assume for the giant peak the influence of local strong scatterers between two ohmic contacts marked in red. (b)~The density of strong scatterers~$n_S$ vs. electron density~$n_e$ for the typical peak (black square) and for the giant peak (red circle) on a log-log scale. (c)~The macroscopic defects on the surface are so called oval defects which are caused by Ga~droplets anywhere in the material. }
\end{figure}

In about $10\,\%$ of all measured contact pairs we observe a giant narrow peak. Figure~\ref{Streuer}~(a) shows the longitudinal resistivity~$\rho_{xx}$ vs. magnetic field around zero magnetic field for two different pairs of ohmic contacts of the same Hall bar. We observe the typical peak~(black) around zero magnetic field  as discussed before and a giant peak~(red) in the same Hall bar. The height and the curvature of the giant peak are clearly different from the typical peak. The strong negative magnetoresistances with a giant peak can also be separated in two distinct sections. The giant peak shows similar dependences on various conditions as before the typical peak, e.~g.  temperature and tilt angle independence. Since we attributed the typical peak to the interplay of smooth disorder with rare strong scatterers, we can conclude that the distribution of the rare strong scatterers seems to be inhomogeneous across the sample.

We get further confirmation of an inhomogeneous distribution by determining the density of strong scatterers~$n_S$ for both types of peaks. Figure~\ref{Streuer}~(b) shows the density of strong scatterers~$n_S$ for the giant peak~(red circles) and for the typical peak~(black squares) vs. electron density~$n_e$. The density of strong scatterers~$n_S$ is determined by using eq.~$\left(\ref{eq:1}\right)$ with the experimental data for the height of the peak~\mbox{$\Delta\rho_{xx}=\rho_0-\rho_{xx}(B_c)$} and its curvature. The density of strong scatterers~$n_S$ as function of electron density is nearly constant for both types of peaks. On the basis of the density~$n_S$ and the mean free path~$\ell_S$ we deduce the average radius of the strong scatteres $a_S=(2\,n_S\,\ell_S)^{-1}\sim19\,\mu$m ($a_S\sim15\,\mu$m for the giant peak). Figure~\ref{Streuer}~(b) also shows that the density of strong scatterers~$n_S$ is higher for the giant peak than for the typical peak. The differences in the density of the strong scatterers~$n_S$ for both types of peaks are signatures for an inhomogeneous distribution of rare strong scatterers between some ohmic contacts.

On the basis of these observations we identify the strong scatterers as macroscopic defects in the material. Our observations fit to the randomly distributed oval defects on the material surface as observed in Fig.~\ref{Streuer}~(c). The low density~$n_S$ and the large radius~$a_S$ also confirm oval defects as strong scatterers. We count on average 28 oval defects in the range of the geometry. The corresponding density of oval defects is $1.3\cdot10^4\,$cm$^{-2}$ which nicely compares with the average density of strong scatterers~$n_S=1.1\cdot10^4\,$cm$^{-2}$ in Fig.~\ref{Streuer}~(b). In some rare cases oval defects apparently 'condense' in some spatial regions of the sample and the giant peak is observed (see the inset in Fig.~\ref{Streuer}~(a)).

Oval defects as seen in Fig.~\ref{Streuer}~(c) arise from the growth process by molecular beam epitaxy. Many proposals concern the formation of oval defects e.~g.~\cite{Chai1981, Pettit1984, Akimoto1985, Weng1986, Shinohara1989}. The common origin of oval defects is attributed to oxides in the Ga~melt. During the growth process Ga~oxides act as nucleation site for unbounded Ga~atoms and Ga~droplets arise. These Ga~droplets lead to locally faster growth of the crystal and cause the formation of oval defects. \mbox{Ga~droplets} occur anywhere in the material. The size of the macroscopic defects observed on our samples varies between a few~$\mu$m and up to 40~$\mu$m which is comparable to the size of strong scatterers as deduced from the peak. Figure~\ref{Streuer}~(c) also shows a schema of the layer structure around the quantum well. \mbox{Ga~droplets} around the quantum well influence the high-mobility 2DEG and are observed as oval defects. The angle independence of the peak (see Fig.~\ref{Winkel}) confirms antidot behavior of oval defects which agrees with the assumption for strong scatterers~\cite{Mirlin2001}.

In conclusion, we have observed a strong negative magnetoresistance at non-quantizing magnetic fields with a peak around zero magnetic field. We have argued that the peak is induced by an interplay of smooth disorder and macroscopic defects while the shoulder next to the peak is dominated by smooth disorder. The macroscopic defects can be observed on the material surface as oval defects.

\begin{acknowledgments}
We would like to thank  \mbox{D. Smirnov} for discussions and for help with the experiments. We are grateful to \mbox{A. P. Dmitriev}, \mbox{A. D. Mirlin} and \mbox{D. G. Polyakov} for discussions and comments. This work was financially supported by the Cluster of Excellence QUEST and by \mbox{DFG-CFN}.
\end{acknowledgments}

\end{document}